\documentclass[aps,pre,twocolumn,superscriptaddress, showkeys ]{revtex4-1}

\usepackage{graphicx,epstopdf}              
\usepackage{subcaption}						
\usepackage{color}
\usepackage{hyperref}                   	
\usepackage{amssymb}               	

\graphicspath{{figs/}}						

\begin{document}

\title{Roughening transition and universality of single step growth models in (2+1)-dimensions}

\author{H. Dashti-Naserabadi}
\affiliation{Physics and Accelerators Research School, NSTRI, AEOI 11365-3486, Tehran, Iran.}
\email{h.dashti82@gmail.com}

\author{A. A. Saberi}
\affiliation{Department of Physics, University of Tehran, 14395-547, Tehran, Iran.}
\affiliation{School of Physics and Accelerators, Institute for research in Fundamental Science (IPM) P.O. 19395-5531, Tehran, 
	Iran.}
\email{ab.saberi@ut.ac.ir}

\author{S. Rouhani}
\affiliation{Department of Physics, Sharif University of Technology 11155-9161, Tehran, Iran.}
\affiliation{School of Physics and Accelerators, Institute for research in Fundamental Science (IPM) P.O. 19395-5531, Tehran, 
	Iran.}

\begin{abstract}
We study (2+1)-dimensional single step model (SSM) for crystal growth including both deposition and evaporation processes parametrized by a single control parameter $p$. Using extensive numerical simulations with a relatively high statistics, we estimate various interface exponents such as roughness, growth and dynamic exponents as well as various geometric and distribution exponents of height clusters and their boundaries (or iso-height lines) as function of $p$. We find that, in contrary to the general belief, there exists a critical value $p_c\approx 0.25$ at which the model undergoes a roughening transition from a rough phase with $p<p_c$ in the Kardar-Parisi-Zhang (KPZ) universality to a smooth phase with $p>p_c$, asymptotically in the Edwards-Wilkinson (EW) class. We validate our conclusion by estimating the effective roughness exponents and their extrapolation to the infinite-size limit.   
\end{abstract}

\keywords{roughening transition, single-step model, universality, scaling exponent}

\maketitle

\section{Introduction}

Roughening transition from a smooth phase with finite width to a rough one
with diverging width is one of the most  
interesting properties of nonequilibrium models for interfacial growth 
\cite{marro2005nonequilibrium, Hinrichsen99Roughening}. A class of nonequilibrium growth processes described by the Kardar-Parisi-Zhang (KPZ) equation \cite{kardar1986dynamic}, is known to be always rough in dimensions $d\le 2$ while exhibits a roughening transition for $d>2$ \cite{Schwartz2012Upper}. The KPZ equation is given by
\begin{equation}\label{eq:KPZ}
\frac{\partial h(\mathbf{x}, t)}{\partial t} = \nu \nabla^2h + \frac{\lambda}{2} \left\vert \nabla h \right\vert ^2 + \eta(\mathbf{x}, t),
\end{equation}
where the relaxation term is caused by a surface tension $ \nu $, and the nonlinear term is due to the lateral growth with strength $\lambda$. The noise $ \eta $ is uncorrelated Gaussian white noise in both space and time with
zero average i.e., $ \langle \eta(\mathbf{x},t) \rangle = 0 $ and $  \langle 
\eta(\mathbf{x},t)\eta(\mathbf{x'},t') \rangle  = 2D\delta^d 
(\mathbf{x}-\mathbf{x'}) \delta(t-t')$. The model produces a self-affine interface $ h(\mathbf{x}) $ whose probability distribution function remains invariant under scale transformation $h(\mathbf{x}) \cong b^{-\alpha} h(b \mathbf{x})$
($\cong$ means statistically the same) with roughness exponent $\alpha \ge 0$.
A possible way to classify various surface growth models is based on scaling behavior of surface width, $w(t,L) = \sqrt{ \langle [ h(\mathbf{x},t)- \langle h \rangle ]^2\rangle }$ where $ \langle \cdot \cdot \rangle $ denotes spacial averaging.
For a nonequilibrium growth surface, the width is expected \cite{family1985scaling} to show the scaling form
$w^2(t,L) \sim L^{2\alpha} f(t/L^z)$, in which
the scaling function $f$ usually has the asymptotic form $f(x \rightarrow \infty) = $ constant and $f(x \rightarrow 0) \sim x^{2\beta}$.
The time $ t_{s} $ when the width first saturates has the scaling ansatz  $ t_{s} \sim L^z $ with the dynamic exponent $z = \alpha/\beta$.  The universality class of a growing interface can then be given by two independent roughness $\alpha$ and growth $\beta$ exponents. For KPZ equation, due to 
additional scaling relation $ \alpha+z=2 $, there remains only one independent exponent, say $ \alpha $ whose exact value is only known in $1d$ \cite{kardar1986dynamic} with $ \alpha=1/2 $. In $ 2d $, the exponent 
is available only by various simulations and theoretical approximations ranging 
from $ \alpha = 0.37 $ to $ 0.4 $ \cite{amar1990numerical, 
marinari2000critical, miranda2008numerical}. Some authors 
\cite{saberi2010classification,saberi2008conformalkpz,
Saberi2008ConformalWo3,saberi2009scaling}
have also argued that it is possible to apply
Schramm-Loewner evolution (SLE) \cite{schramm2000scaling} based on statistics and fractal properties of iso-height lines of saturated 2$d$ surfaces to 
classify surface growth processes as well.

In $d>2$, there exists a critical value $\lambda_c$ for the nonlinearity coefficient in Eq. \ref{eq:KPZ} which separates flat and rough surface phases \cite{imbrie1988diffusion,tang1990multicritical}. In the weak coupling (flat) regime ($\lambda<\lambda_c$) the nonlinear
term is irrelevant and the behavior is governed by the
$\lambda=0$ fixed point i.e., the linear Edwards-Wilkinson (EW)
equation \cite{edwards1982surface} whose exact solution is known: $\alpha = (2-d)/2$ and $z=2$. In the more challenging
strong-coupling (rough) regime ($\lambda>\lambda_c$), where the nonlinear
term is relevant, the behavior of the KPZ equation
is quite controversial and characterized by anomalous exponents.
There is, however, a longstanding controversy (see e.g., \cite{Schwartz2012Upper} and \cite{SaberiEPL2013} and references therein)
concerning the existence and the value of an upper critical
dimension $d_c$ above which, regardless of the strength
of the nonlinearity, the surface remains flat. The aim of this paper is to investigate the possibility of roughening transition and universality of 2$d$ single step discrete growth model (SSM) which, to our best knowledge, has not been addressed before. A coarse-graining derivation of the SSM surface dynamics in (1+1)-dimensions has revealed \cite{PrivmanBook2005} that it belongs to the KPZ universality class. Although there is no rigor theoretical support for this claim in higher dimensions, it is believed to be true in any spatial dimension $d>1$ as well. However, our study can shed light on the controversial relation between SSM and KPZ model as well as the roughening transition of the KPZ equation in (2+1)-dimensions.

Various discrete models have been suggested in the past to describe surface growth processes (see e.g., \cite{barabasi1995fractal,meakin1997fractals}). Among them, here we study the class of $ 2d $ single step models  
(SSM) \cite{meakin1986ballistic, plischke1987time, liu1988universality, 
kondev2000nonlinear}, a kind of solid on solid (SOS) 
models \cite{kim1989growth} 
which is defined as follows: the growth starts 
from an initial condition $h(i,j;t=0) =  [ 1+(-1)^{i+j}]/2 $ with $ 1\le i 
\le L_x $ and $ 1\le j \le L_y $, on a square lattice of size $L_x\times L_y$. At each step one site $ (i,j) $ is randomly chosen, if $h(i,j)$ is 
a local minimum then it is increased by $ 2 $ with probability $p_+$ 
(deposition process), and if it is a local maximum then  its 
height is decreased by $ 2 $ with probability $p_-$ (desorption 
or evaporation process). Such definition guarantees that at each step, the 
height difference  between two neighboring sites would be exactly 1. 
Overhanging is 
not 
allowed in this model and the interface will not develop large slopes. Without loss of generality, we consider $p_+ + 
p_-=1$ that leaves only one control parameter $p:=p_+ \le 0.5$ (up-down symmetry switches $p_+\leftrightarrow 1-p_+$ ) which is believed to play the same role as the nonlinearity coefficient in the KPZ equation \ref{eq:KPZ} as $\lambda\leftrightarrow(p-0.5) $.

This model has been investigated in the past, claiming that for $ p=0.5 $ and $ p \ne 0.5 $, it belongs to the EW and KPZ universality classes, respectively \cite{plischke1987time, liu1988universality, kondev2000nonlinear}.
Plischke \textit{et al.} \cite{plischke1987time} have shown that for $ p=0.5 $ 
in $ 1d $, this model is reversible and can be exactly solved by mapping to the 
kinetic Ising model. They have found $ \alpha = 1/2 $ and $ z=2 $. Furthermore, 
for $ p \ne 0.5 $ they have mapped the interface model onto the driven hard-core lattice gas, and focused on the average slope of the interface. In an approximate way, they have then shown that the equation of the average slope is in agreement with the Burgers's equation \cite{plischke1987time}, thus claiming that the universality class is that of KPZ equation for  $ p \ne 0.5 $.  They have also simulated this model for $ p=0.25 $, and claimed that in the limit of large system sizes $ L $, the exponent $ z $ converges to $ z^{\mathrm{KPZ}} = 3/2 $ in 1$d$. Simulations by the same authors on SSM in $ 2d $ \cite{liu1988universality}, have provided the scaling exponents $ \alpha \approx 0 $, $ z \approx 2 $ for $ p=0.5 $, and $ \alpha \approx  0.375$, $ z \approx 1.64 $ for $ p = 0 $. 
Kondev \textit{et al.} \cite{kondev2000nonlinear}, have also simulated SSM on 
a square lattice of size $ L=128 $, and confirmed that the model for $ p=0.5 $ 
and $ p=0.1 $ are consistent with the EW and KPZ classes, respectively. 
However, 
they found 
that $ p=0.3 $ consistently resembles $ p=0.5 $, contrary to the claims 
in \cite{plischke1987time}, and they attributed their finding to a slow crossover from initially Gaussian to asymptotic KPZ behavior.
A generalized single step model has also been investigated in \cite{gates1988growth,gates1988kinetics,gates1988kinetics2}   which exhibits a dynamical crossover characterized by a shift in the early-time scaling exponent from its KPZ value to the EW value. This has been first explained in \cite{krug1992amplitude,krug1990mechanism} by showing that this behavior is due to a change in the sign of the nonlinear parameter $\lambda$. It is also known that the (2+1)-dimensional anisotropic KPZ equation with lambdas of opposite sign does generate EW, rather than KPZ scaling behavior \cite{halpin1992kinetic,wolf1991kinetic}.

In this paper we are going to revisit the model in (2+1)-dimensions and present the results of  extensive simulations with relatively large 
system sizes and higher precision. We will estimate various geometrical exponents as function of the control parameter $p$ in the two following sections \ref{exponents} and \ref{clusters}. We will estimate the roughness exponent by extrapolating the results to the infinite-size limit in Sec. \ref{effective} and come to the conclusion in Sec. \ref{concl} that there exists a critical value $p_c$ at which SSM exhibits a roughening transition from a rough phase with $p<0.25$ to a smooth phase with $p>0.25$.

\section{Interface exponents}{\label{exponents}}
\label{sec:surfexponents}

In this section we present the first part of our results obtained from extensive simulations on a square lattice of size $ 50 \le L \le 700$, in which the averages for $w(t,L)$ are taken over more than $200$ independent runs. We estimate the roughness $\alpha$ and dynamic $z$ exponents by examining the scaling laws i.e., $w_s\sim L^\alpha$ (where $w_s$ is the saturated width) and $t_s\sim L^z$, respectively. We use the system size $L=4000$ to estimate the growth exponent $\beta$ by using the scaling ansatz $w(t)\sim t^\beta$ for $t<t_s$.
 To compute various geometric exponents of iso-height lines and height clusters in the next sections, the averages are taken over $ 10^4 $ height configurations on a square lattice of size $ L=1000$. To further justify our conclusion, we perform simulations on a rectangular geometry of size $ L_x = 3L_y $ and $ L_y = L $ with $ 100 \le L  \le 1000$ to measure the winding angle statistics of the iso-height lines and their fractal dimensions. One time step is defined as $L^2$ number of trials for particle deposition or evaporation.

\begin{figure}
 \includegraphics[width=0.45\textwidth]{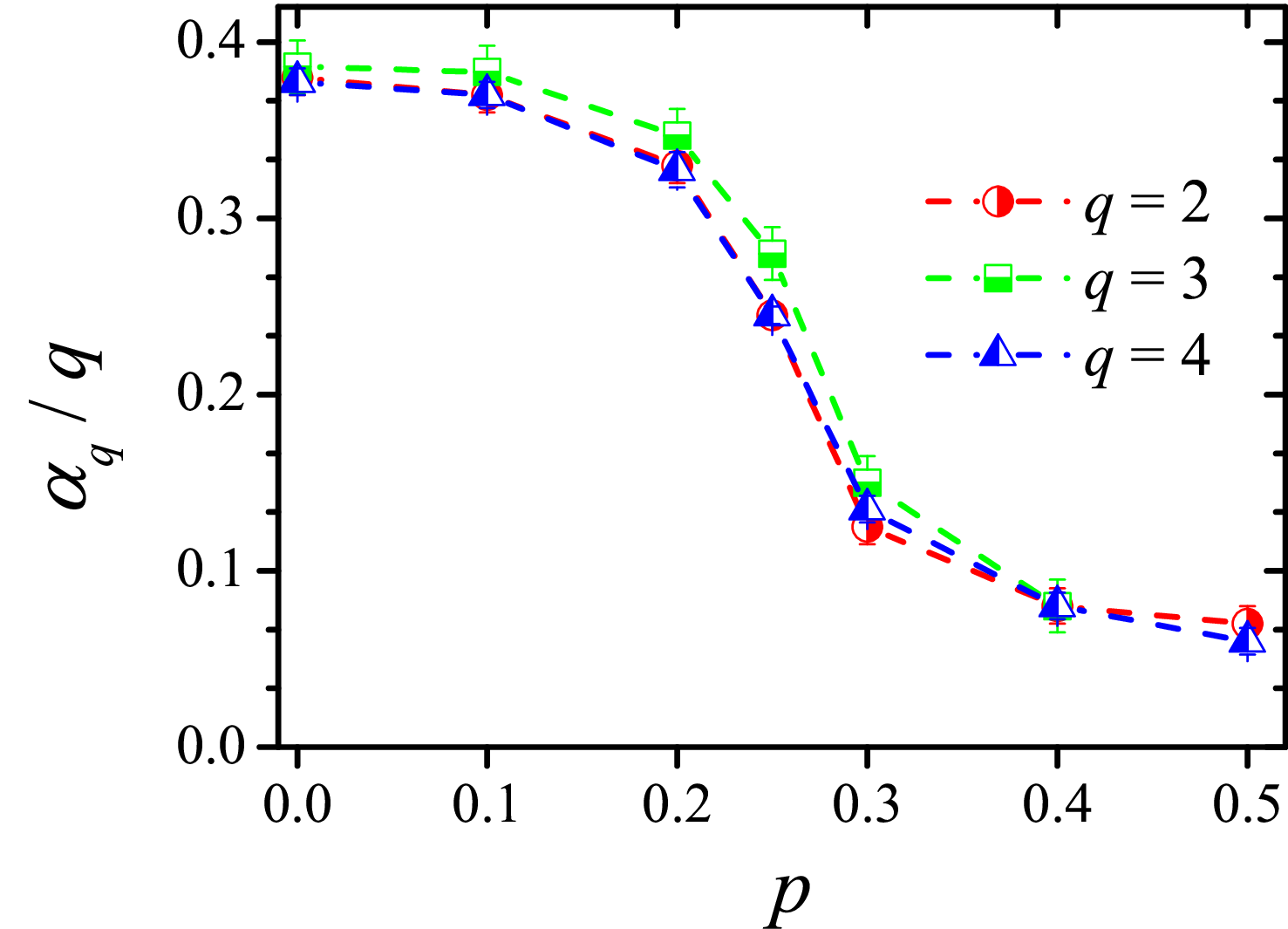}
 \caption{The rescaled exponent $\alpha_q/q$ as a function of the control parameter $p$ for different values of $q=1, 2, 3$. For self-affine surfaces $\alpha_q/q$ has to be independent of $q$ which is the case here, within the error bars, for the SSM grown interfaces.}
 \label{fig:a_q234-L1000}
 \end{figure}

To check the efficacy of our simulations, let us first estimate the roughness exponent $\alpha$ from \textit{scale-dependent curvature} in the saturation regime $t>t_s$. The curvature $ C_b(\mathbf{x}) $ at position $ \mathbf{x} $ on scale $ b $  is defined as follows \cite{kondev2000nonlinear}
 \begin{equation}\label{eq:cb_x}
 C_b(\mathbf{x}) = \sum_{m=1}^{M} \left[  h(\mathbf{x} + b\mathbf{e}_m)- h(\mathbf{x}) \right],
 \end{equation}
where the offset directions $ \{\mathbf{e}_m\}_{m=1}^M $ are a fixed set of vectors summing up to zero. In our case on a square lattice, $ \{\mathbf{e}_m\} $  are pointing along the $ \{10\} $ type directions.
For a self-affine surface, the curvature is expected to satisfy the following scaling  relation \cite{kondev2000nonlinear}:
 \begin{equation}\label{eq:cb}
 \left< C_b(x)^q \right> \sim b^{\alpha_q}  \quad  \mathrm{with} \  \alpha_q = q\alpha ,
 \end{equation}
where $\langle \cdot \cdot \rangle$ denotes spatial averaging.
To check this relation, simulations are carried out on square lattice of size  $L=10^3$ with more than $ 10^4 $ number of height configurations for different values of $p$. We apply periodic boundary conditions in both directions. The estimated rescaled exponents $\alpha_q/q$ for three different values of $ q $  are plotted in Fig. \ref{fig:a_q234-L1000} as a function of the control parameter $p$.  We find that the three curves for $q=2,3$ and $4$ are independent of $q$ (within the error bars), confirming the self-affinity of the height profiles. Since the curvature vanishes at $p=0.5$ for $ q=3$, the point is excluded in the plot. 
The other important feature observed in Fig. \ref{fig:a_q234-L1000} is the crossover between two KPZ and EW universality classes with $\alpha\approx 0.38$ and $\alpha\approx 0$, respectively. This is the main goal of the present study to clarify if there is a roughening transition at a critical control parameter $p_c \ne 0$ in the sufficiently large system size limit in 2$d$.

\begin{figure}
 \includegraphics[width=0.45\textwidth]{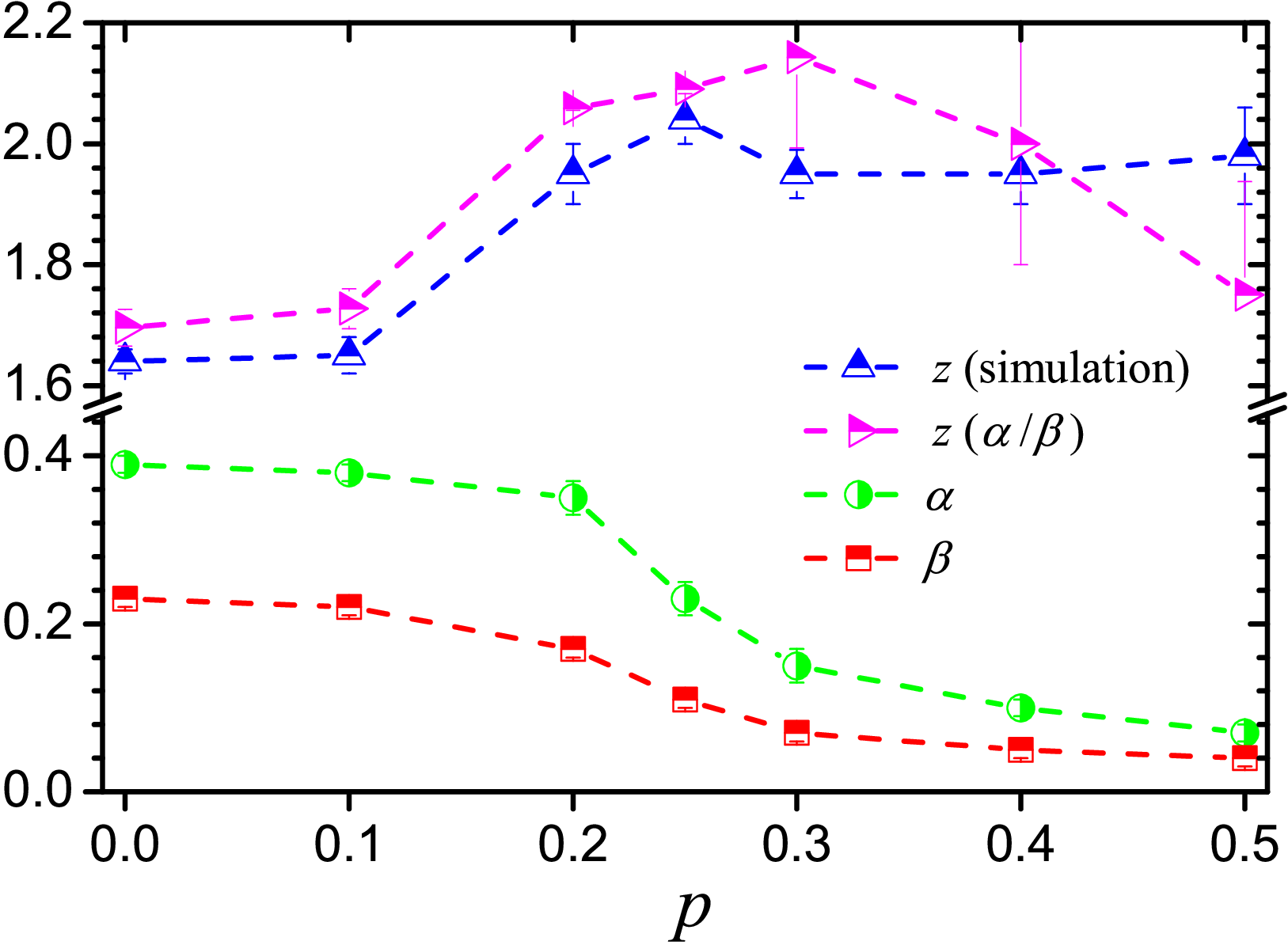}
 \caption{(Color online): Interface exponents including roughness $\alpha$, growth $\beta$ and dynamic $z$ exponents as function of $p$. The dynamic exponent is computed directly from the scaling relation $ t_s \sim L^ z$ (up-triangles) and by using the relation $z=\alpha/\beta$ (right-triangles). The rather high error bars for the latter are caused by the fact that both the roughness $\alpha$ and growth $\beta$ exponents vanish for $p>0.25$.} 
\label{fig:abz-several-p}
 \end{figure}

The interface exponents $\alpha, \beta$ and $z$ as function of $p$ are shown in Fig. \ref{fig:abz-several-p} (each exponent is measured independently). For the two limiting cases at $p=0$ and $p=0.5$, the exponents are again in good agreement with those for KPZ and EW universality classes, respectively \cite{amar1990numerical,marinari2000critical, miranda2008numerical}. However, we find that except for the intermediate values around $p_c\approx 0.25$, the plots suggest that the exponents are approximately equal within the two disjoint intervals $p<p_c$ and $p>p_c$. This observation can be a benchmark of roughening transition at $p_c\approx 0.25$ which calls for a more delicate analysis. In the following sections we present various observations of different geometric exponents as function of $p$ which confirm our observation. We will then justify our finding by estimating effective exponents and extrapolations to the infinite system size.

\section{Statistics of the height clusters and iso-height lines}{\label{clusters}
\label{sec:statloops}

In this section we present the results of our further analysis on the fractal properties of the height clusters and iso-height lines as well as the scaling properties of the distribution of the cluster size and their perimeter. We find that the corresponding exponents show characteristic behavior below and above $p_c\approx 0.25$, unraveling further information about the self-affinity of the interfaces \cite{kondev2000nonlinear, kondev1995geometrical}.

Consider an ensemble of height configurations in the saturated regime. For each configuration, a cut is made at a specific height $h_\delta = \langle h \rangle + \delta \sqrt{\langle  [ h(x) - \langle h \rangle ]^2\rangle}:=0$, where $\delta$ is a small real number indicating the level of the cut. Each island (or cluster height) is defined  as a set of nearest neighbor sites with positive height identified by the Hoshen-Kopelman algorithm \cite{hoshen1976percolation}. Let us first consider $\delta = 0$, i.e., the cut is made at the average height level. The iso-height lines (or loops) can be uniquely determined by the algorithm explained in \cite{saberi2009thermal}. In order to illustrate how islands behave as function of $p$, the snapshots of the positive height clusters are shown in Fig. \ref{fig:iso-height-clusters} for $p=0, 0.1, 0.2, 0.3, 0.4$ and $0.5$. 
As it is evident in the figure, the islands are more compact for lower $ p $, and become more porous and scattered for larger $p$. This picture is also in agreement with the cluster geometries previously observed \cite{saberi2008conformalkpz} for KPZ and EW models.

\begin{figure*}[]
	\centering
	\begin{subfigure}{0.2\textwidth}\includegraphics[width=\textwidth]{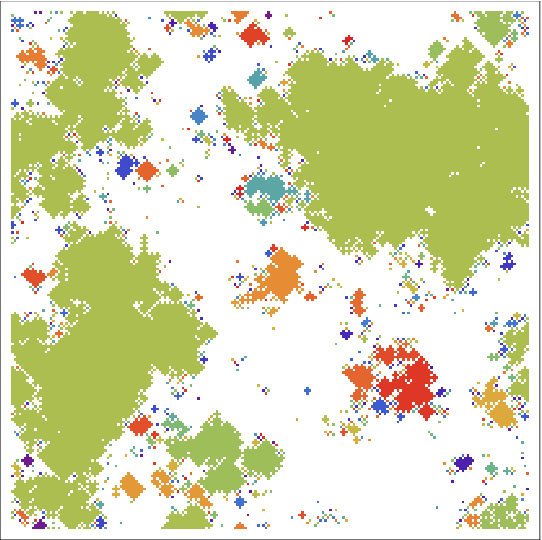}\caption*{$p = 0$}
    \end{subfigure}
    \begin{subfigure}{0.2\textwidth}\includegraphics[width=\textwidth]{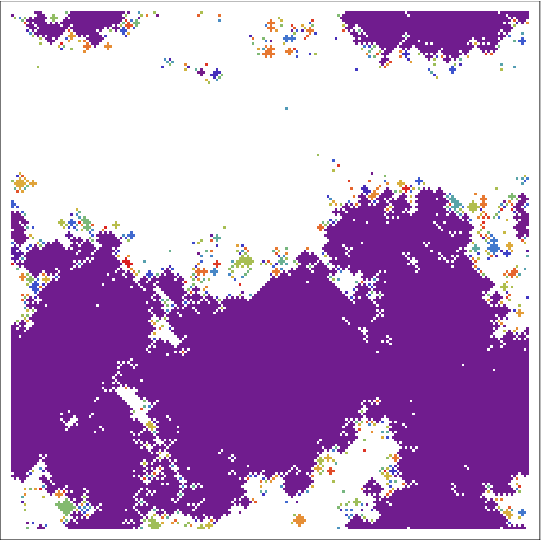}\caption*{$p = 0.1$}
    \end{subfigure}
    \begin{subfigure}{0.2\textwidth}\includegraphics[width=\textwidth]{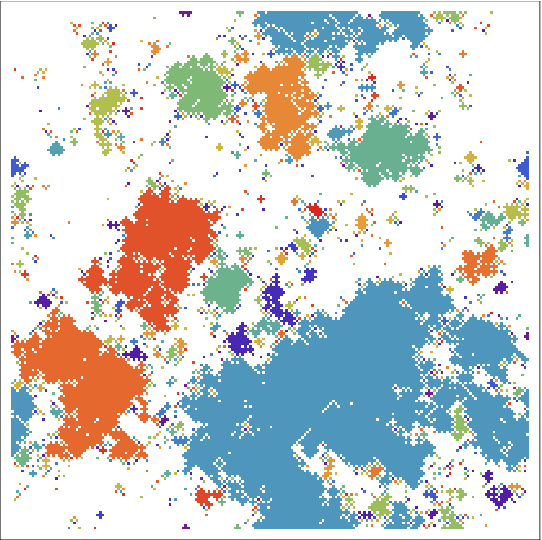}\caption*{$p = 0.2$}
    \end{subfigure}

    \begin{subfigure}{0.2\textwidth}\includegraphics[width=\textwidth]{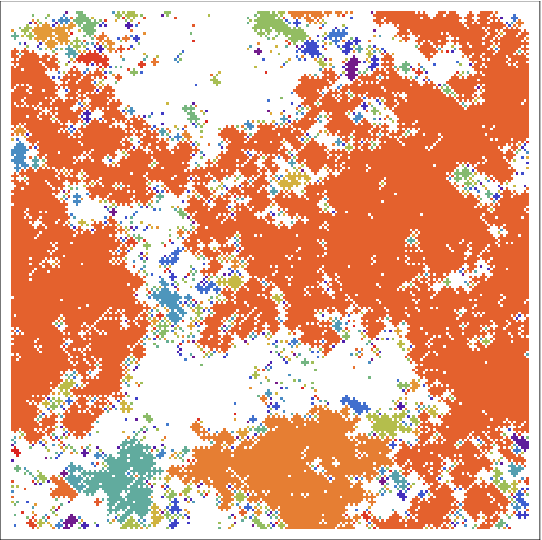}\caption*{$p = 0.3$}
    \end{subfigure}
	\begin{subfigure}{0.2\textwidth}\includegraphics[width=\textwidth]{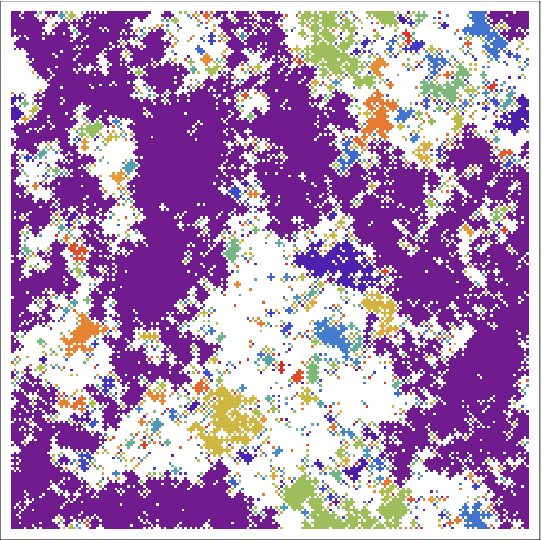}\caption*{$p = 0.4$}
    \end{subfigure}
    \begin{subfigure}{0.2\textwidth}\includegraphics[width=\textwidth]{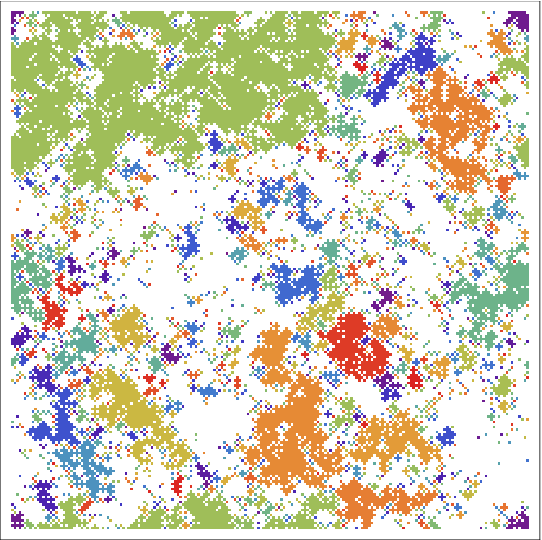}\caption*{$p = 0.5$}
    \end{subfigure}

  \caption{(Color online): Snapshots of positive height clusters for different values of $ p $ on a square lattice of size $ L=200 $. The cut is made at the average height.   }
  \label{fig:iso-height-clusters}
 \end{figure*}

\textit{\textbf{Fractal dimensions.}}
Self-similarity of the clusters offers a scaling relation between the average mass $M$ of a cluster and its radius of gyration $R$,  i.e., $M \sim R^{D_c}$, with $D_c$ being the fractal dimension of clusters. The average length $l$ of a cluster boundary also scales with it's radius of gyration $r$ as $l \sim  r^{d_f}$ \cite{kondev2000nonlinear,kondev1995geometrical}. Moreover, the relation between the average area $a$ of a loop and it's perimeter is given by $l \sim a^{d_a}$ where $d_a = d_f/2$ (for compact clusters).
To estimate these fractal dimensions, we generate more than $ 10^4 $ samples of height configurations on a square lattice of size $ L=1000 $. As an example, we present the data for the scaling of $ l(r) $ in Fig. \ref{fig:l_r-severalp} for various values of $p$, whose slope in the log-log scale gives the corresponding fractal dimension. The estimated exponents are reported in Fig. \ref{fig:df-L1000}.
\begin{figure}
 \includegraphics[width=0.45\textwidth]{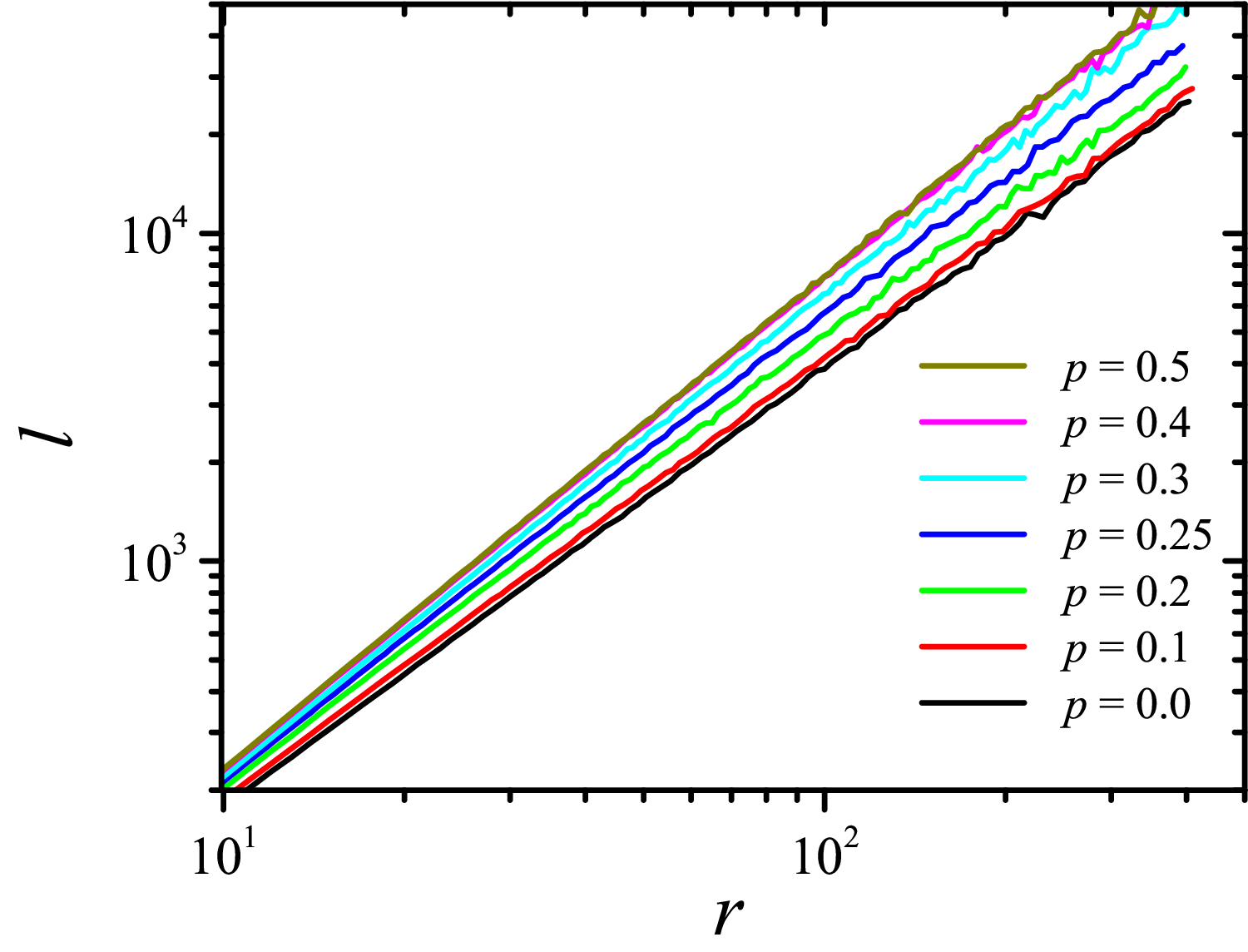}
 \caption{(Color online): The average length of a cluster boundary versus the average radius of gyration on square lattice of size $ L =1000 $. Averages are taken over more than $ 10^4 $ height configurations.}
 \label{fig:l_r-severalp}
\end{figure}
\begin{figure}
 \includegraphics[width=0.44\textwidth]{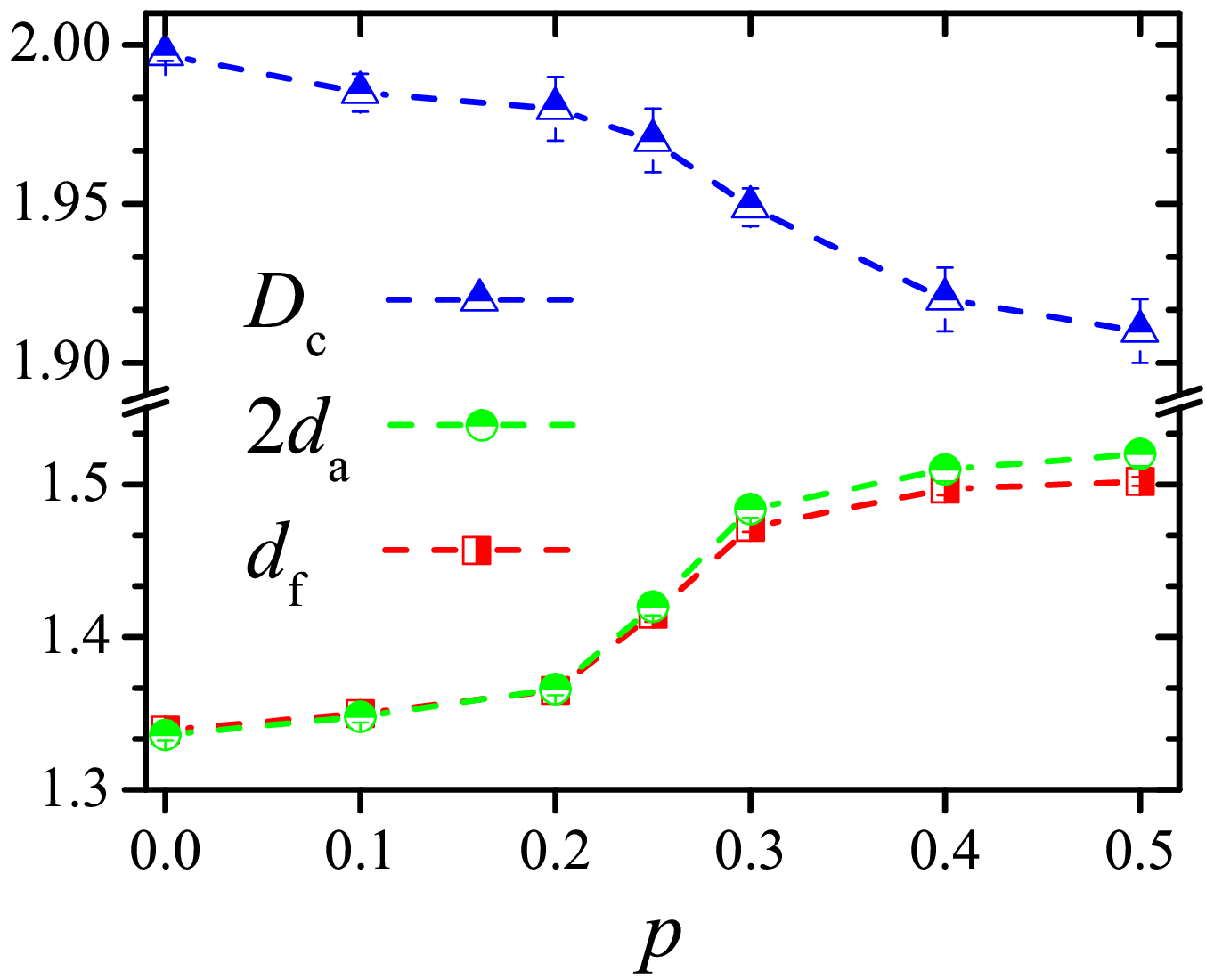}
 \caption{(Color online): Fractal dimensions of the height clusters and their boundaries as function of $ p $.}
 \label{fig:df-L1000}
\end{figure}
We find that all these fractal dimensions cross over between two limiting KPZ and EW classes \cite{saberi2008conformalkpz,saberi2009scaling, schramm2010contour}. We have also checked that the exponents do not depend on the level $\delta$ of the cut, although the range of scaling slightly does \cite{saberi2010geometrical}. In order to see the finite-size effects, we have also measured the exponents by going to the larger sizes up to $ L=3000 $ with a number of $ 2500 $ independent samples, and found similar results within the error bars.

\begin{figure}
 \includegraphics[width=0.45\textwidth]{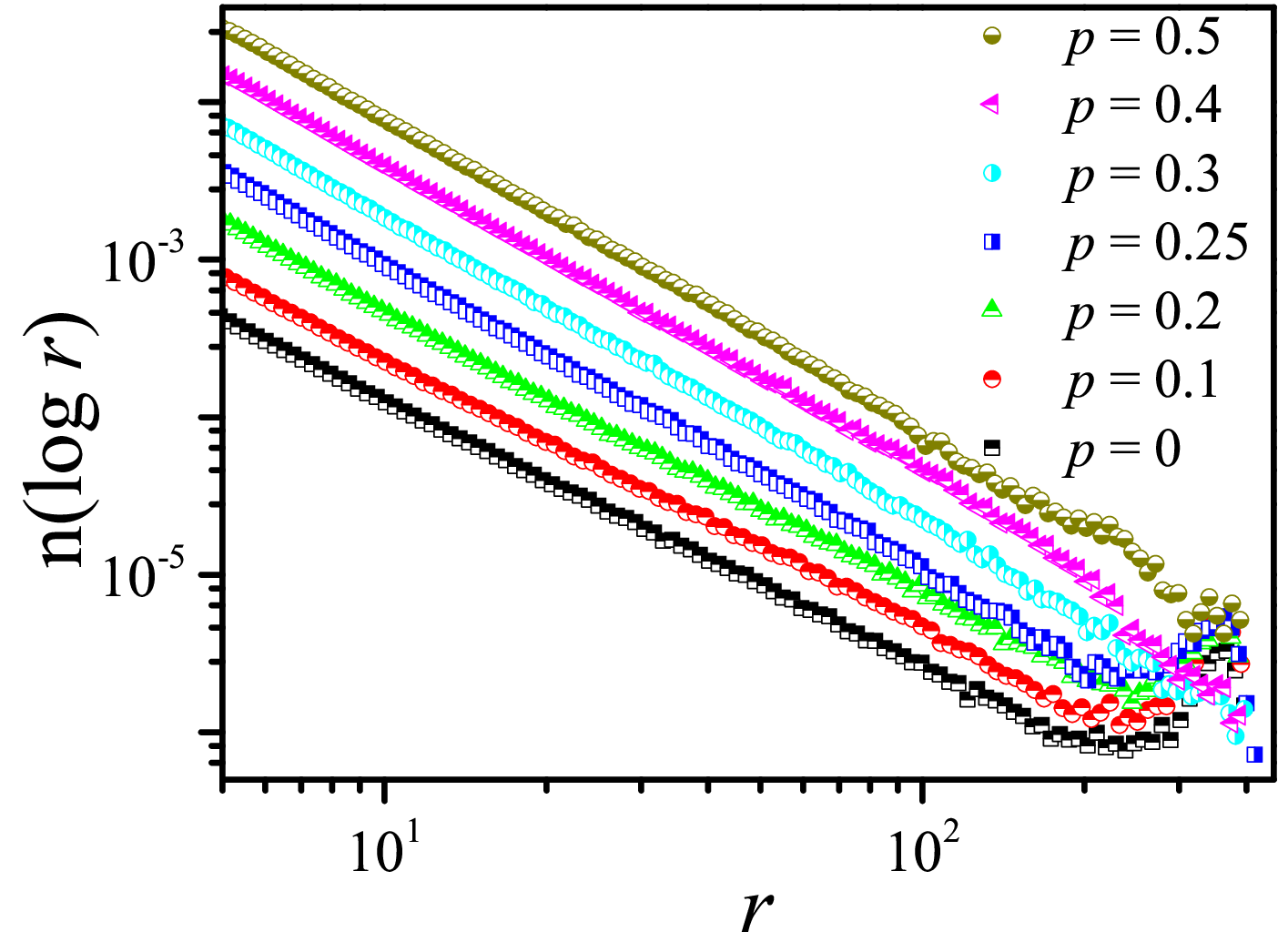}
 \caption{(Color online): Distribution function of $ \log r $  versus $ r $ (the average radius of gyration) on square lattice of size $ L=1000 $, for various values of $ p $. The slope gives the exponent $ \tau_r-1 $ shown in Fig. \ref{fig:tau-L1000}. For more clarity, the plots are suitably shifted.}
 \label{fig:n_r-severalp}
\end{figure}
\begin{figure}
 \includegraphics[width=0.45\textwidth]{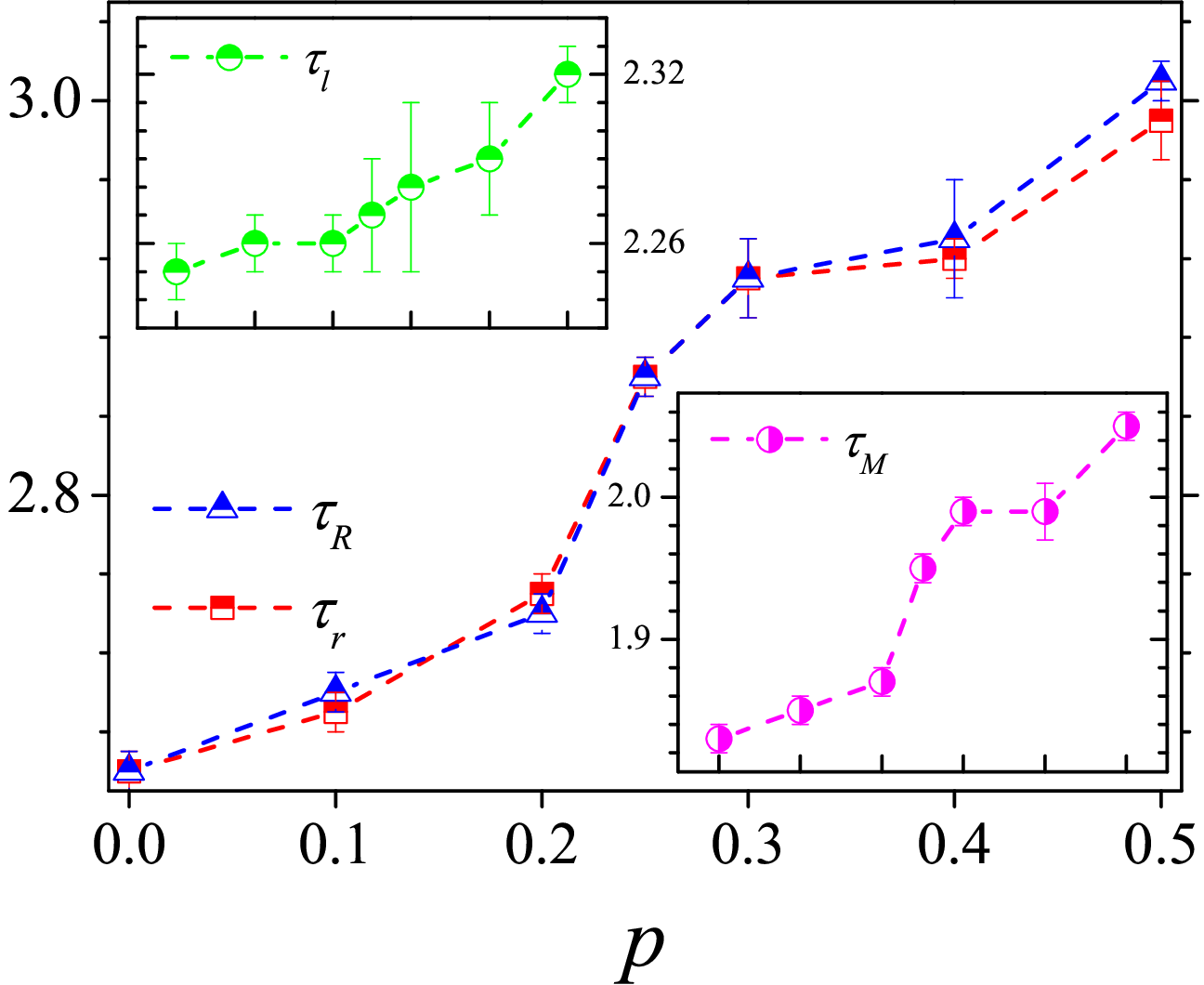}
 \caption{(Color online): Various distribution exponents (see the text) as function of $ p $.}
 \label{fig:tau-L1000}
\end{figure}

\textit{\textbf{Distribution exponents.}}
We now look at the distribution functions of different statistical observables of the height clusters and contours, such as the contour length distribution $ n(l) $, cluster size distribution $ n(M) $ and distributions for the radius of gyration of the contours $n(r)$ and clusters $n(R)$. We confirm that all these distributions follow the scaling forms i.e., $ n(l) \sim l^{-\tau_l} $, $ n(M) \sim M^{-\tau_M} $, $n(r) \sim r^{-\tau_r}  $ and $ n(R) \sim R^{-\tau_R}  $ \cite{kondev2000nonlinear,kondev1995geometrical} (see Fig. \ref{fig:n_r-severalp} for an example).
All distribution exponents are summarized in Fig. \ref{fig:tau-L1000} for $\delta=0$ as function of $p$ which again confirm the crossover behavior. The exponents $ \tau_R $ and $ \tau_r $ coincide within the error bars. In the following, we investigate dependence of the distribution exponents on the level $\delta$ of the cut as previously noted by Olami \textit{et al}. \cite{olami1996scaling}

\begin{figure}
 \includegraphics[width=0.45\textwidth]{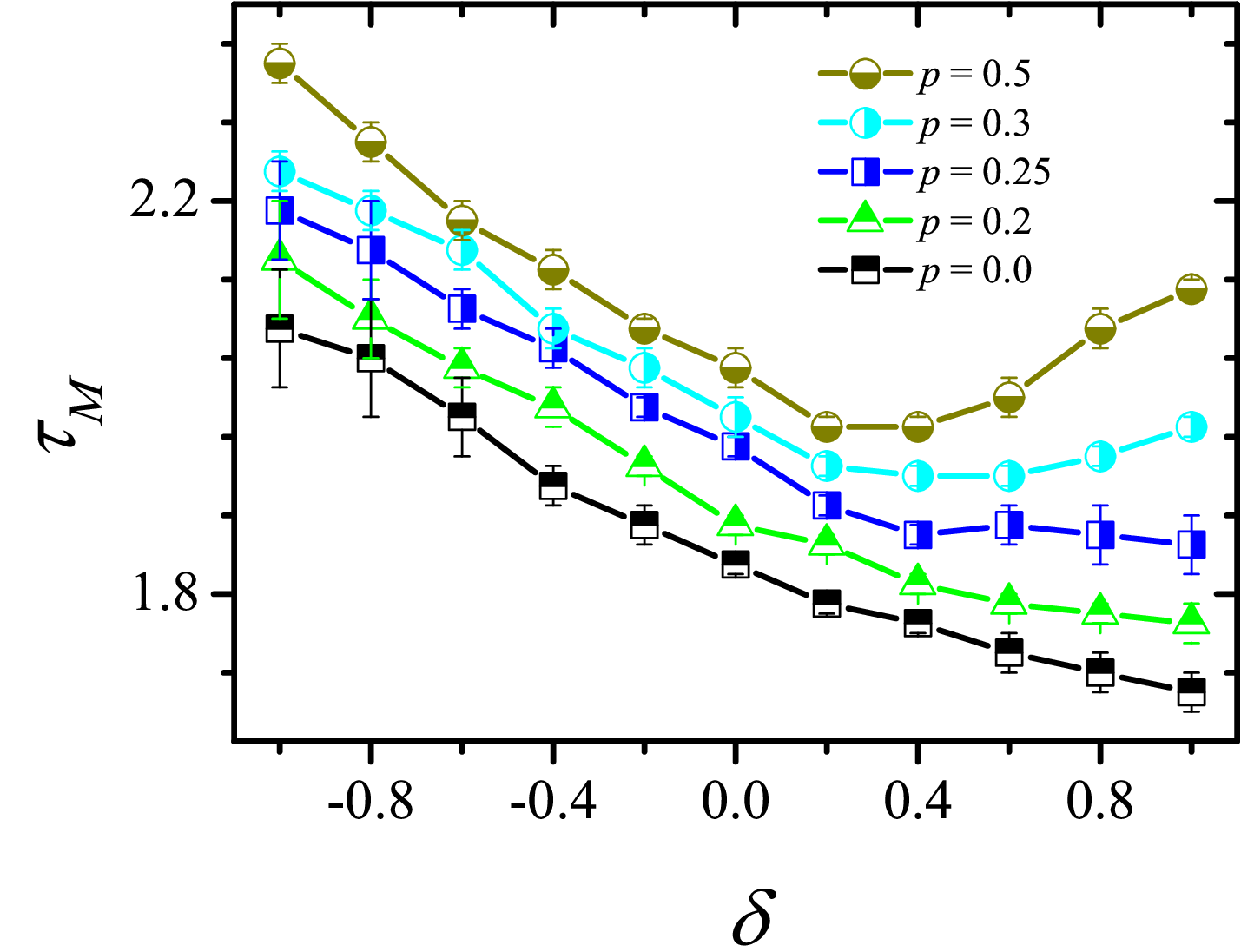}
 \caption{(Color online): The island-size distribution exponent as a function of the level cut $\delta$ for various $ p $. }
 \label{fig:tau_M-delta}
\end{figure}
\begin{figure}
 \includegraphics[width=0.45\textwidth]{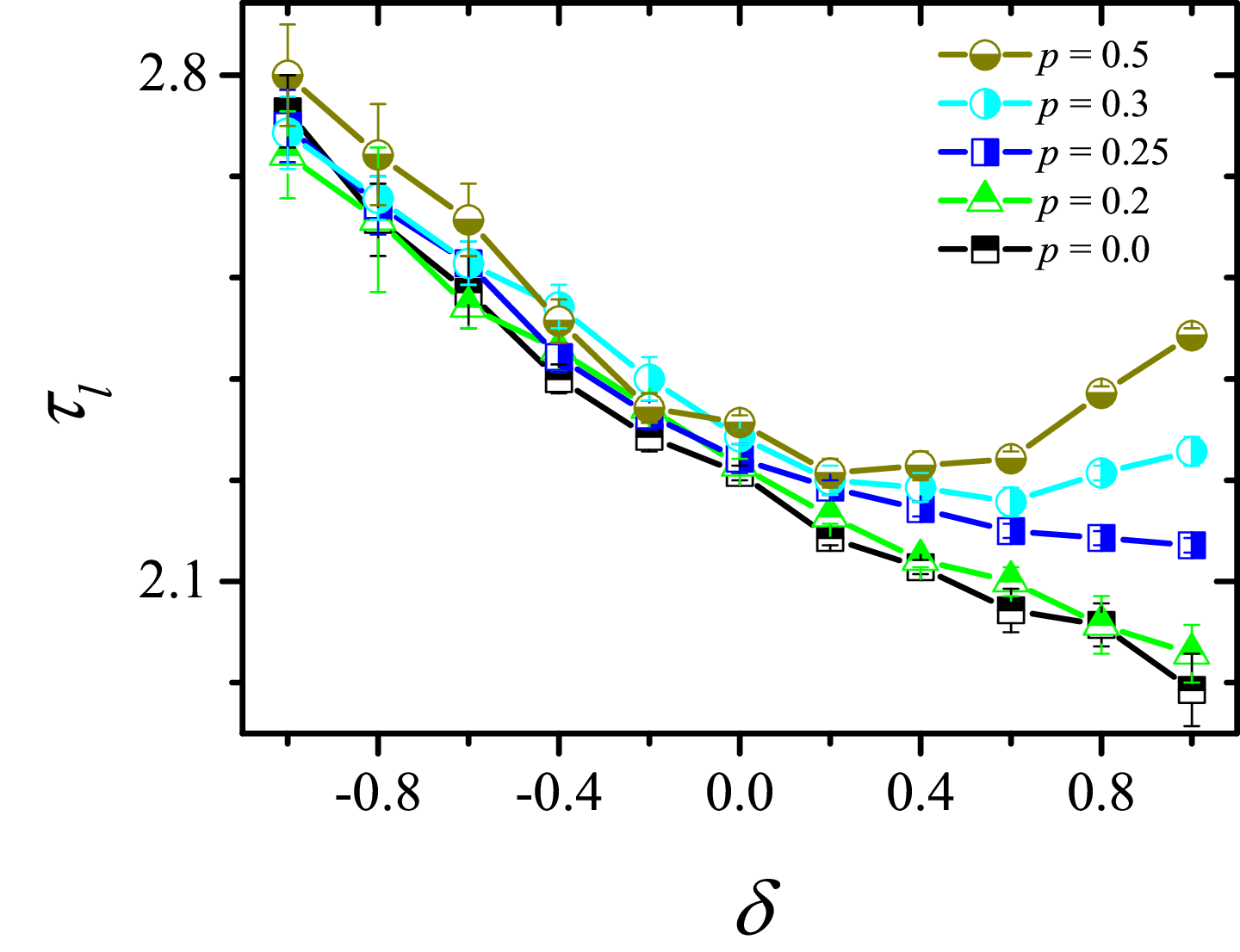}
 \caption{(Color online): The distribution exponent for the length of the height cluster boundaries as a function of the level cut $\delta$ for various $ p $. }
 \label{fig:tau_l-delta}
\end{figure}

\textit{\textbf{Dependence of the exponents on $\delta$.}}
All previous results were obtained at the mean height level i.e., at $ \delta=0  $. Let us now examine their dependence on the level of the cut. Our analysis reveal that the fractal exponents such as fractal dimension of contours $d_f$ and clusters $D_c$, do not depend on $\delta$. Nevertheless, our results show that the distribution exponents do change with $\delta$. As shown in figures \ref{fig:tau_M-delta} and \ref{fig:tau_l-delta}, the exponents show a bowl-like functionality to $ \delta $ for $p>0.25$ while for $p<0.25$, they monotonically decrease with $ \delta $.

\textit{\textbf{Winding angle statistics.}} Here we present the results of independent extensive simulations of SSM on a strip geometry of size $ L_x \times L_y $ with $ L_x=3L_y $ and $ L_y=L $. For each height configuration, we find all spanning clusters at level $ \delta=0 $ in $ y $ direction, and assign corresponding coastlines that connect the lower boundary to the upper one.
We consider $ L = 100, 200, 300, 400, 500$ and $1000 $, and examine the scaling relation $ l \sim L^{d_f} $, to compute the fractal dimension $d_f$ of the  spanning curves (Fig. \ref{fig:l_L-severalp}). We could gather a number of $ 10^4 $ spanning curves from an approximately $ 7500 $ independent saturated height profiles. 
\begin{figure}
\includegraphics[width=0.45\textwidth]{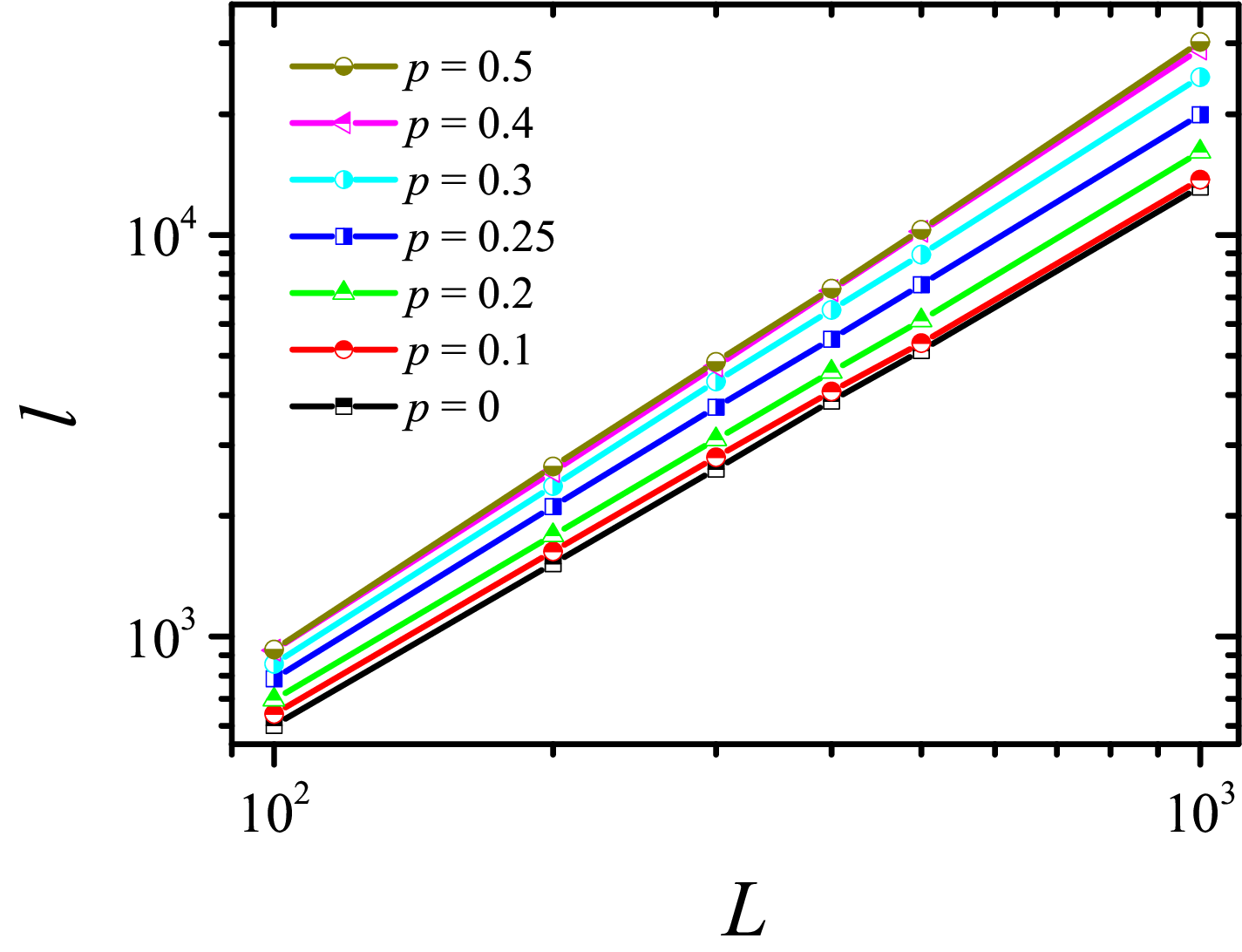}
\caption{(Color online): The average length $l$ of a spanning curve on a strip geometry versus the width $ L_y=L $ of the strip. The slopes give the fractal dimension of the corresponding iso-height line for various $p $. }
\label{fig:l_L-severalp}
\end{figure}

We compute the winding angle $ \theta $ of the curves as defined by Wieland and Wilson \cite{wieland2003winding}. For each curve we attribute an arbitrary winding angle to the first edge (that is set to be zero). The winding angle for the next edge is then defined as the sum of the winding angle of the present edge and the turning angle to the new edge measured in radians. The variance of the winding angle is believed to behave like $ \langle \theta^2 \rangle \sim a + b \ln L $ \cite{wieland2003winding}, where for conformal curves $ b=2(d_f-1) $. We have computed the variance of the winding angle for an ensemble of spanning iso-height curves for different $ p $ as function of lattice size $ L $, and confirmed that is linearly proportional to its logarithm  with a universal coefficient $ b $ which depends on $ p $ (see Fig. \ref{fig:theta2_L-severalp}). The two computed fractal dimensions from direct measurement ($l \sim L^{d_f} $) and $ d_f=b/2+1 $, are plotted in Fig. \ref{fig:df_p-sevrealL} for a comparison. They almost coincide for $ p>0.25 $ but slightly deviate for $ p<0.25 $. They both however present a crossover behavior around $p_c\approx 0.25$.
\begin{figure}
 \includegraphics[width=0.45\textwidth]{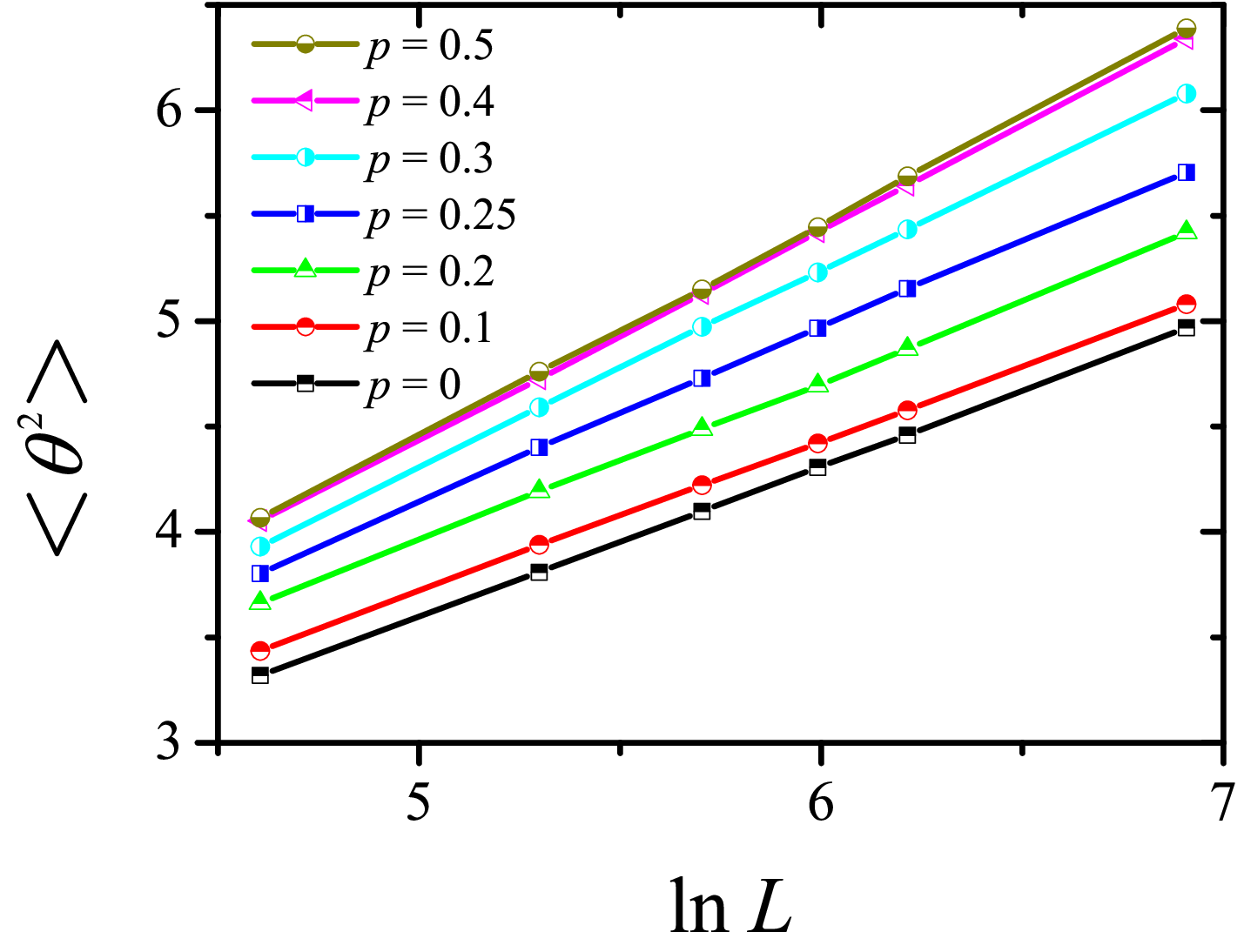}
 \caption{(Color online):  The variance of winding angle versus logarithm of the lattice width for various $ p $. }
 \label{fig:theta2_L-severalp}
\end{figure}
\begin{figure}
 \includegraphics[width=0.45\textwidth]{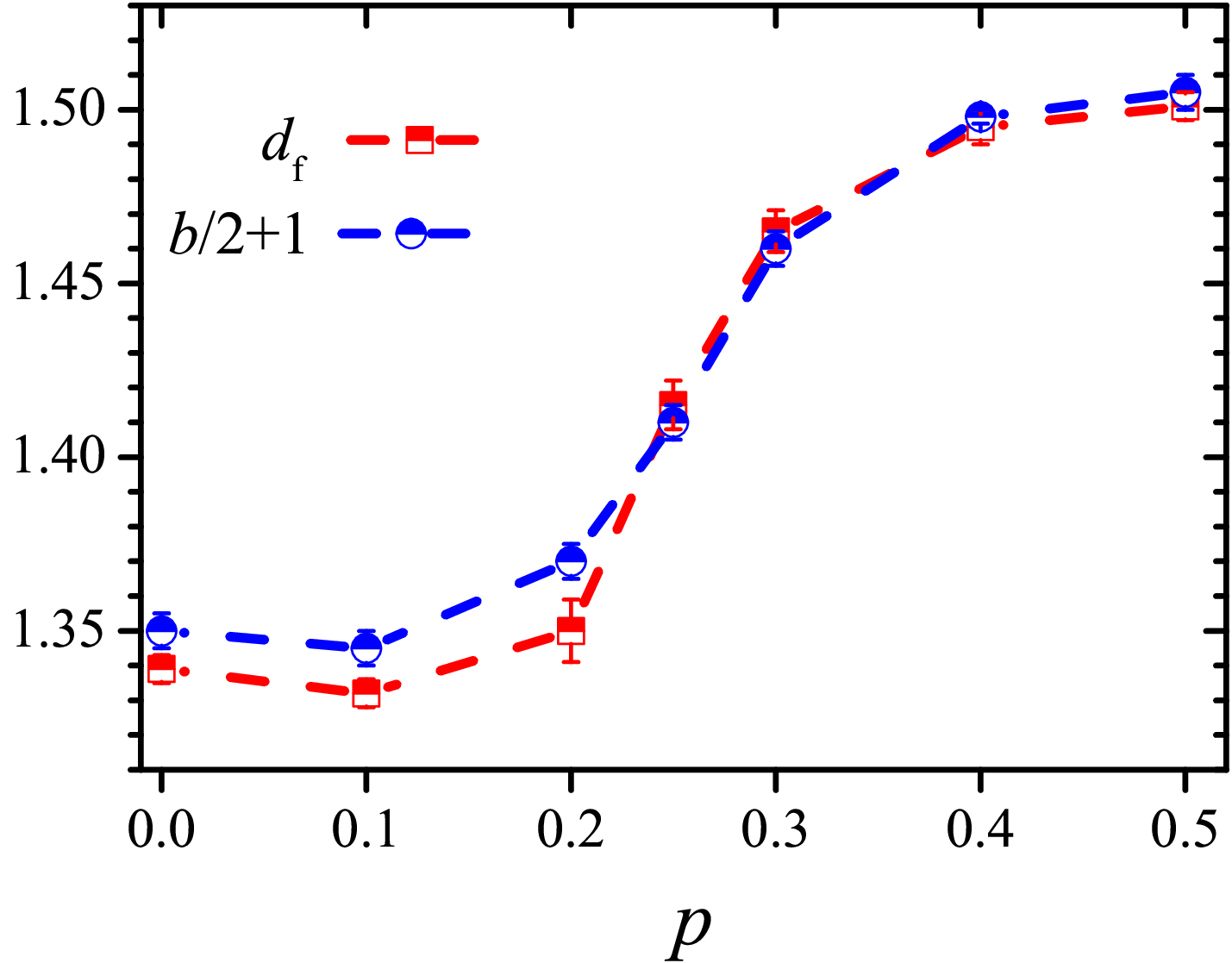}
 \caption{(Color online):  The fractal dimensions obtained from the scaling relation $ l \sim L^d_f $ (squares) compared with the one derived from the slopes of the linear plots in Fig. \ref{fig:theta2_L-severalp} (circles). }
 \label{fig:df_p-sevrealL}
\end{figure}
\begin{figure}
\includegraphics[width=0.45\textwidth]{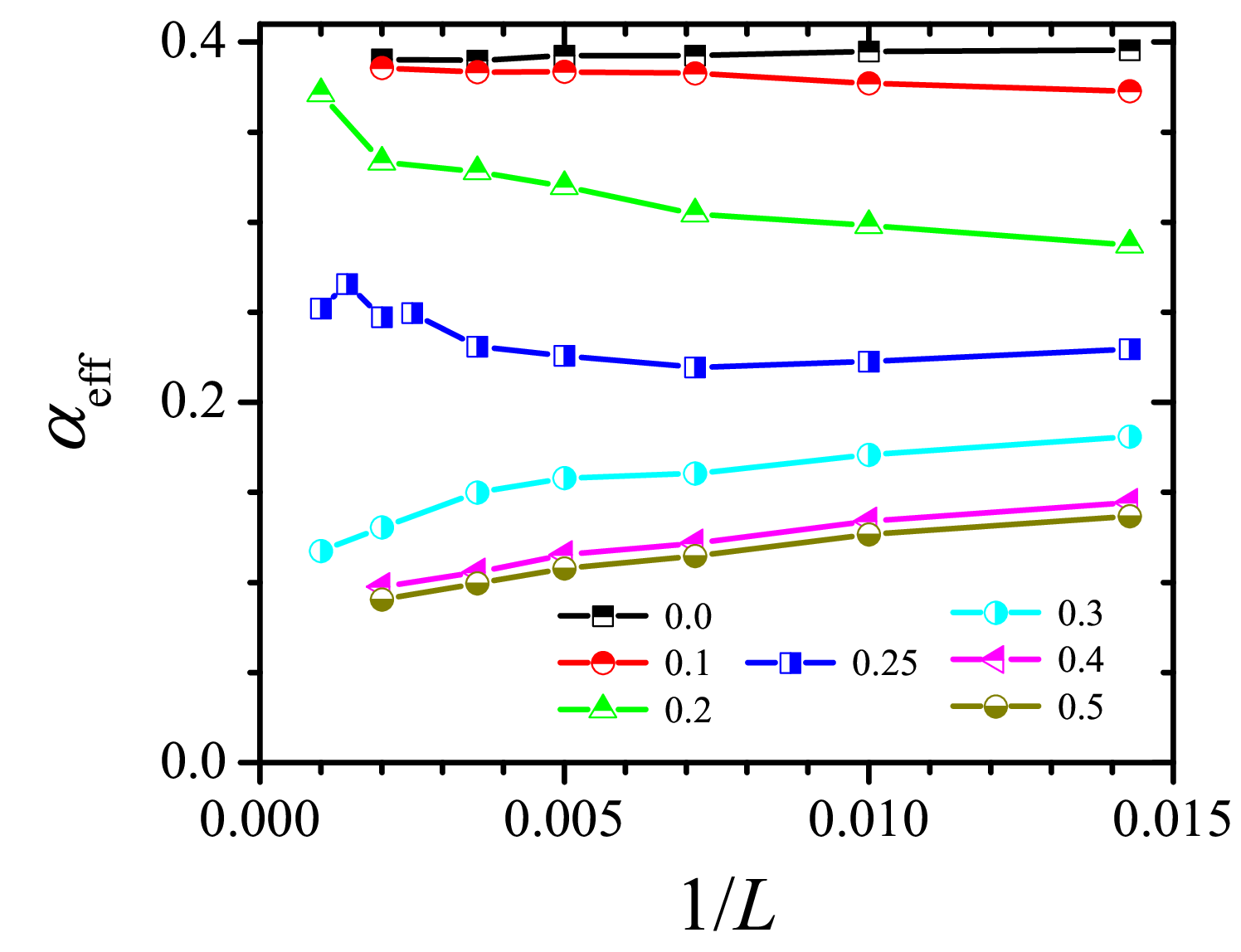}
\caption{(Color online): The effective roughness exponent $\alpha_{\mathrm{eff}}$ as function of $1/L$ for several values of $p$. The error bars are less than $10^{-3}$. For $p<0.25$ and $p>0.25$, $\alpha_{\mathrm{eff}}$ converges to $\alpha^{\mathrm{KPZ}}\approx 0.38 $ and $\alpha^{\mathrm{EW}}\approx 0$, respectively.}
\label{fig:a_eff}
\end{figure}

\section{Effective Exponents }{\label{effective}

In the previous section we have shown that various geometric exponents have a crossover behavior between two limiting KPZ and EW classes which seems to approach a sharp step-like roughening transition around $p_c\approx 0.25$ in the thermodynamic limit. Although we have used relatively large system sizes with adequate statistics in our computations, there may however exist systematic deviations from the true thermodynamic values. In order to eliminate the systematic errors, we compute size-dependent effective roughness exponent \cite{KIM2012PRE} for various values of $p$.

The effective roughness exponent $\alpha_{\mathrm{eff}}(L_k)$, is defined by the successive slopes of the line segments connecting two neighboring points of $( L_{k-1},w_s(L_{k-1}) )$ and $( L_{k},w_s(L_{k}) )$ in which $w_s(L_k)$ stands for the saturated width for an SSM grown interface on a square lattice of size $L_k$ averaged over more than $2\times 10^3$ independent runs,
\begin{equation}\label{eq:d_f-eff}
\alpha_{\mathrm{eff}}(L_k)=\frac{\ln[w_s(L_k)/w_s(L_{k-1})]}{\ln[L_k/L_{k-1}]}.
\end{equation}
It is plotted against $1/L$ in Fig. \ref{fig:a_eff} to extrapolate the roughness exponent in the
infinite-size limit. We find that the roughness exponents for $p<0.25$ converge to the known KPZ
roughness exponent $\approx 0.38$ and for $p>0.25$ asymptotically converge to the EW value $\approx 0$ in the limit $L\rightarrow\infty$.

\begin{figure}
\includegraphics[width=0.46\textwidth]{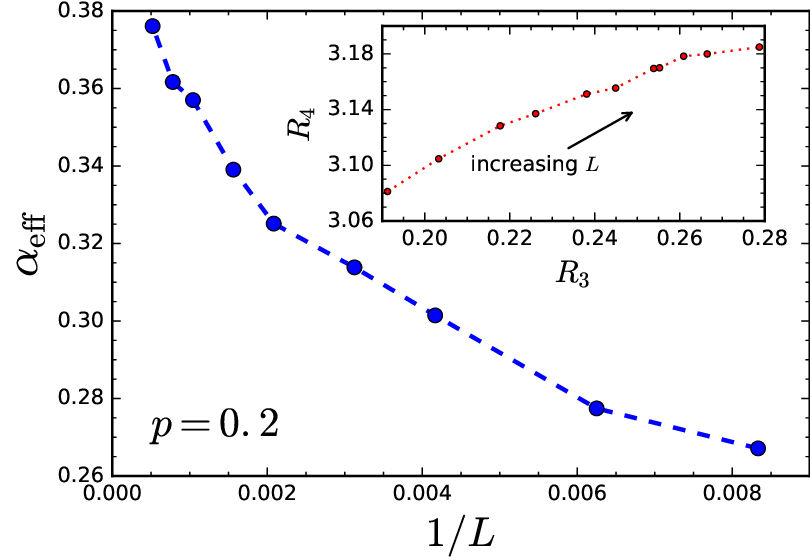}
\includegraphics[width=0.45\textwidth]{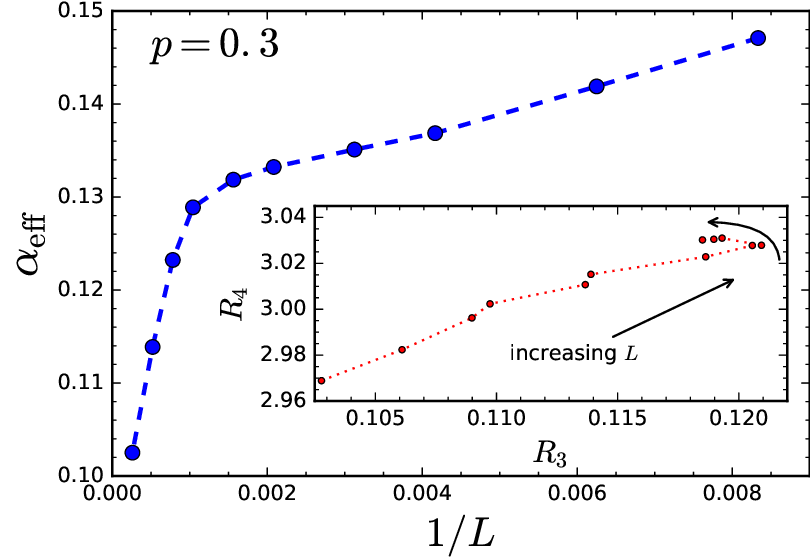}
\caption{(Color online): The effective roughness exponent $\alpha_{\mathrm{eff}}$ as defined in Eq. \ref{eq:alpha-eff} as a function of $1/L$ for two boundary values $p=0.2$ (Main top) and $p=0.3$ (Main bottom) around the critical value $p_c\approx 0.25$. The error bars are less than $10^{-3}$. In the Insets the cumulant of kurtosis $R_4$ is presented as a function of the skewness $R_3$ for various sizes. The statistics of the corresponding grown surfaces for $p<0.25$ and $p>0.25$ converges to the KPZ and EW universality classes, respectively.}
\label{fig:a_eff-2}
\end{figure}

\begin{figure}

\includegraphics[width=0.46\textwidth]{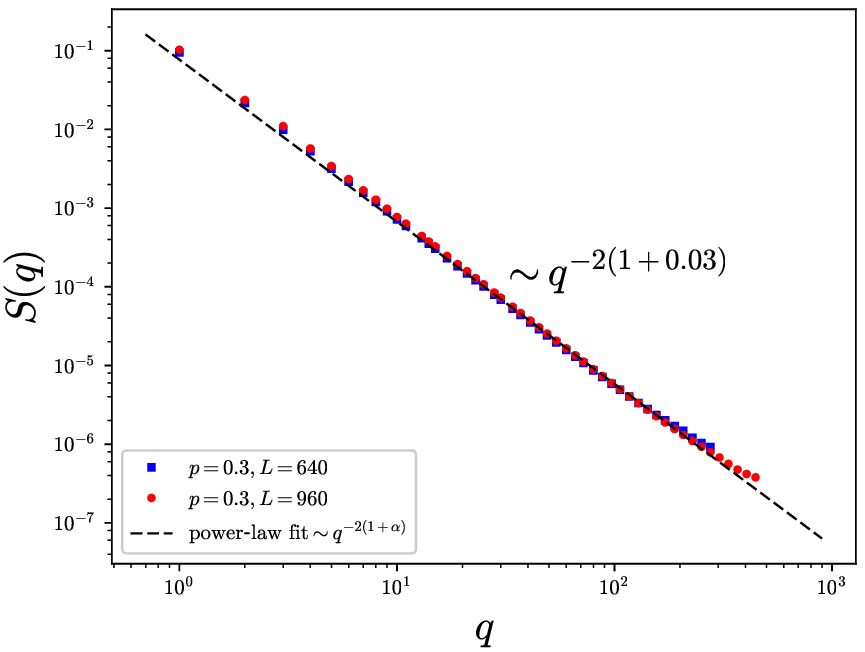}
\caption{(Color online): Height structure factor $S(q)$ as a function of $q$ for $p=0.3$. The comparison between the scaling ansatz $S( \mathbf{q})\sim | \mathbf{q}|^{-2(1+\alpha)}$ and the best fit to our data (dashed line) gives $\alpha=0.03(3)$, in a good agreement with the EW universality class. }
\label{fig:S_q}
\end{figure}

In order to further strengthen our conclusion on existence of a roughening transition around $p_c\approx 0.25$, let us now focus our attention on two boundary values i.e., $p=0.3$ and $p=0.2$ around $p_c$. We follow the analysis presented in \cite{parisi2015numerical} in which a careful finite-size scaling analysis of the critical
exponents, and an accurate estimate of the first three moments of the height fluctuations, are used to estimate the roughness exponent of the Restricted Solid on Solid model in $d=2$ with a rather high accuracy. To this aim, we run independent extensive simulations of the SSM on square lattices of various linear size 
$L=60$, $80$, $120$, $160$, $240$, $320$, $480$, $640$, $960$, $1280$, $1920$, $3840$ for $p=0.3$ (which is a more challenging case), and also all sizes except $L=3840$, for $p=0.2$. For each $L$ and $p$ we generate more than $10^4$ samples for averaging.
Statistical sampling is adopted at steady state regime ($t>t_s$). For a given sample at time $t>t_s$, we measure the first three
connected moments $w_n(L,t) = (1/L^2)\sum_{i=1}^{L^2}(h_i(t)- \bar{h})^n$ where $\bar{h}=(1/L^2)\sum_{i=1}^{L^2}h_i(t)$ and $n=2, 3, 4$. Then we define the asymptotic (in time) estimate as
$w_n(L) = \frac{1}{T+1} \sum_{t=t_s}^{t_s+T} w_n(L,t)$ for $T\gg t_s$.

To appreciate more clearly the finite-size effects on $\alpha$, we evaluate the effective roughness exponent $\alpha_\mathrm{eff}$ with a slight modification of Eq. \ref{eq:d_f-eff} \cite{parisi2015numerical} as
\begin{equation}\label{eq:alpha-eff}
\alpha_{\mathrm{eff}}(L) = \frac{\log \left( w_2(L)/w_2(L') \right) }{2\log(L/L')},
\end{equation}
where $L/L'=2$. We also compute the cumulants of skewness $R_3=w_3/w_2^{3/2}$ and kurtosis $R_4=w_4/w_2^2$. For the Gaussian (EW) surfaces these quantities are known to be $R_3=0$ and $R_4=3$.

Figure \ref{fig:a_eff-2} summarizes the results of our computations for $p=0.2$ (top) and $p=0.3$ (bottom).
For $p=0.2$, $\alpha_\mathrm{eff}$ clearly approaches to that of the KPZ universality class. The ratio of the cumulants $R_4$ versus $R_3$ is also plotted in the Insets of Fig. \ref{fig:a_eff-2}. For $p=0.2$ a significant departure from a normal distributed
fluctuation of the surface is observed. \\
In contrast to the observed behavior for $p=0.2$, our data for $p=0.3$ strongly supports our previous conclusion that the SSM for $p>0.25$ belongs to the EW universality class, as displayed in Fig. \ref{fig:a_eff-2} (bottom). The effective roughness exponent asymptotically converges to that of the EW class in the limit $L\rightarrow\infty$. As shown in the Inset, the ratio of the cumulants $R_4$ versus $R_3$ are more consistent with a normal distribution where we find $R_4=3.00(3)$. Although $R_3$ increases for small system sizes but it starts decreasing for larger $L$ (note the direction of arrows for the increasing system size). 

In the context of surface kinetic roughening, a very important quantity is the two-dimensional height structure factor i.e., $S(\mathbf{q})=\langle |\tilde{h}(\mathbf{q})|^2\rangle$, where $\tilde{h}(\mathbf{q})$ is the space Fourier transform of $h(\mathbf{x})-\bar{h}$. This function has many advantages over real-space correlation
functions, specially in the presence of crossover behavior and anomalous scaling \cite{korutcheva2004advances} where is frequently shown to be less affected by crossover effects. As a final and independent cross-check, we have carried out simulations for $p=0.3$ (which is more controversial) of sizes $L=640$ and $960$ to compute $S( \mathbf{q})$ and estimate the corresponding roughness exponent from its scaling behavior i.e., $S( \mathbf{q})\sim | \mathbf{q}|^{-2(1+\alpha)}$ \cite{kondev2000nonlinear}. \\
As displayed in Fig. \ref{fig:S_q}, we find the roughness exponent $\alpha=0.03(3)$ for $p=0.3$ which is, to a good extent, in agreement with the EW universality class.\\Therefore, all our computations indicate that there is an unexpected roughening transition for single step growth models in (2+1)-dimensions around $p_c\approx 0.25$.

 It is worth mentioning that in the context of the related problem of directed polymers in random media (DPRM), member of the KPZ universality class, Imbrie and Spencer \cite{imbrie1988diffusion} have provided a rigorous mathematical proof that the model in (2+1)-dimensions, as in (1+1)-dimensions, is strictly strong-coupling and super-diffusive ($z<2$), except at the isolated point of infinite temperature, where the wandering is simply entropic (i.e., $z=2$), analog of the EW stochastic growth behavior. For higher dimensions i.e., transverse substrate dimensions of $d=2+\epsilon$, a finite-temperature roughening transition does exist, but for $\epsilon=0$, there is a complicated multi-critical behavior involving very long, exponentially divergent time scales. This has been studied in an impressive series of works \cite{tang1990multicritical,forrest1990surface,tang1992kinetic}  on  ($d+1$)-dimensional hypercubic-stacking (HCS) models with $d=1,2$ and $3$, in which the authors show that a nonequilibrium surface-roughening transition occurs in $d=3$, but in $d=2$ they have only observed a smooth crossover behavior rather than a true roughening transition. The reason for this discrepancy may be as follows. HCS model and SSM are identical only in $d=1$, and for $d>1$ the microscopic growth rules are different since the height difference of neighboring
columns in HCS model becomes $1$ and $-d$. This imposes additional up/down asymmetry in favor of the KPZ fixed point which delays the asymptotic convergence and thus the observation of a true roughening transition in the parameter space. 
It is intriguing that a more careful look at the presented data in Fig. 11 of \cite{tang1992kinetic} for (2+1)-dimensional HCS simulations shows a real compression of the effective exponents for $p>0.25$ which may be the signature of a roughening transition in the asymptotic limit. In order to verify this postulated asymptotic convergence in $2d$ SSM, we carried out new simulations to produce the same data as in the Fig. 11 of \cite{tang1992kinetic} for $2d$ SSM of rather large sizes up to $L=2^{14}$ for $p=0.35$ and $L=2^{13}$ for other values of $p$. As shown in Fig. \ref{fig:w2t-wt}, our data confirms again the existence of a roughening transition around $p\approx 0.25$.

\begin{figure}
\includegraphics[width=0.48\textwidth]{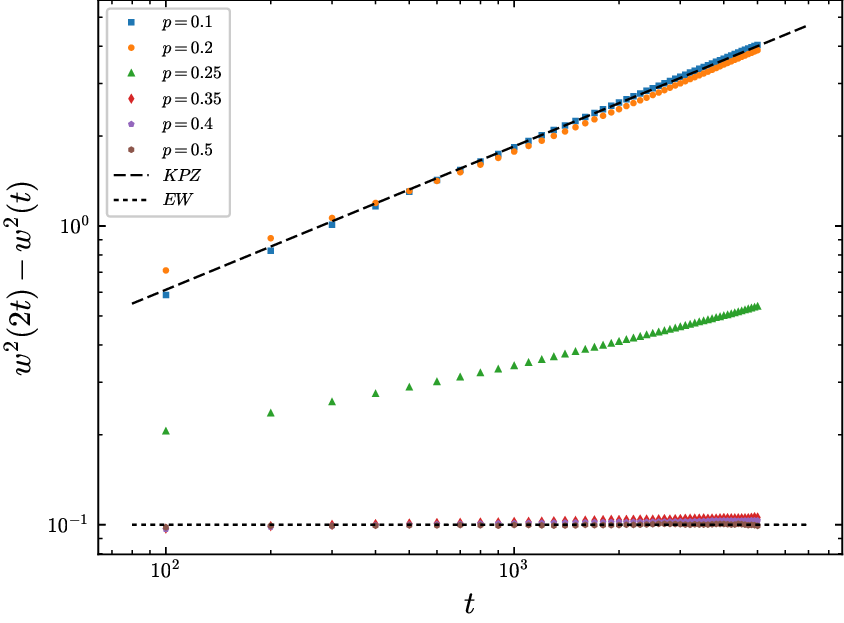}
\caption{(Color online): Surface width data vs time for various values of $p$ around $p=0.25$, compared with the KPZ (longer-dashed line) and the EW scaling (shorter-dashed line). The data are shifted appropriately upwards for clarity. }
\label{fig:w2t-wt}
\end{figure}

However, the discrepancy with \cite{imbrie1988diffusion} which establishes the marginality of $d=2$ case, could arise from some peculiarity of the microscopic growth rules of these discrete growth models in (2+1)-dimensions whose delicate understanding will be the line of our future research.

There also exist some known results that can additionally be tested, which will be the purpose of our future work. Most notably, it is known that (2+1)-dimensional KPZ interfaces display one-point height
fluctuations described by a (generalized) Tracy-Widom probability distribution function \cite{halpin20122dim, oliveira2013KPZ,halpin2013extermal}, which should hold for $p<0.25$ and be falsified for $p>0.25$.

\section{Conclusion}
\label{concl}

We have studied the single step model (SSM) for crystal growth in (2+1)-dimensions which admits both deposition and evaporation processes parametrized by a single control parameter $0\le p\le 0.5$. There is a general consensus that the model belongs to the KPZ universality class for $p=0$ and EW class for $p=0.5$. However, various studies in the past have considered the control parameter $(p-0.5)$ proportional to the nonlinearity coefficient $\lambda$ in the KPZ equation \ref{eq:KPZ} and concluded that the model asymptotically belongs to the KPZ universality class for all $p\ne 0$.  

In this paper we have presented the results of extensive simulations and obtained satisfactory evidence which rule out the previous claims. Extrapolations to the infinite-size limit reveal that there exists a critical value $p_c\approx 0.25$ around which the model exhibits a roughening transition from a rough phase with $p<0.25$ in the KPZ universality to the asymptotically smooth phase with $p>0.25$ in the EW universality class.

Our study opens a new stimulating challenge in the field and calls for further theoretical investigations of the model. An interesting question arises concerning the upper critical dimension $d_u$ of the model and its relation to the same controversial problem in the KPZ model which is the main subject of our future work. However, according to the previous studies \cite{halpin1990disorder} and  \cite{forrest1990surface,tang1992kinetic} on HCS model, the upper critical dimension should be $d_u>3$.

\section*{Acknowledgement}
A.A.S. would like to acknowledge supports from the Alexander von Humboldt Foundation, and partial financial supports by the research council of the University of Tehran.


\bibliography{refs}

\end{document}